# Cold, dry, windy, and UV irradiated – surveying Mars relevant conditions in Ojos del Salado Volcano (Andes Mountains, Chile)

Kereszturi A.[1,10], Aszalós J.[2], Heiling Zs.[3], Ignéczi Á.[9], Kapui Zs.[4], Király Cs.[5], Leél-Össy Sz.[6], Nagy B.[7], Nemerkényi Zs.[5], Pál B.[1], Skultéti A.[5], Szalai Z.[5,8]

[1] Konkoly Thege Miklos Astronomical Institute, Research Centre for Astronomy and Earth Sciences, H-1121 Konkoly-Thege Miklos 15-17, Budapest, Hungary.

[2] Department of Microbiology, Eötvös Loránd University, 1117 Budapest Pázmány Péter 1/C, Hungary.

[3] Földgömb Foundation for Research Expeditions, H-1142 Budapest, Erzsébet királyné útja 125., Hungary.

[4] Research Centre for Astronomy and Earth Sciences, Institute for Geological and Geochemical Research, H-1112 Budapest, Budaörsi út 45.

[5] Geographical Institute, Research Centre for Astronomy and Earth Sciences, H-1112 Budapest, Budaörsi út 45.

[6] Department of Physical and Applied Geology, ELTE Eötvös Loránd University, 1117 Budapest Pázmány Péter 1/C, Hungary.

[7] Department of Physical Geography, ELTE Eötvös Loránd University, 1117 Budapest Pázmány Péter 1/C, Hungary.

[8] Department of Environmental and Landscape Geography, Eötvös Loránd University, H-1117 Budapest, Pázmány Péter 1/C, Hungary.

[9] Department of Geography, University of Sheffield.

[10] European Astrobiology Institute, 1, quai Lezay-Marnésia - BP 90015, 67080 Strasbourg Cedex, France

## Abstract

The Special Collection of papers in this issue of *Astrobiology* provide an overview of the characteristics and potential for future exploration of the Ojos del Salado volcano, located in the Andes Mountains in front of the Atacama Desert in northern Chile. The main benefits of this site compared with others are the combination of strong UV radiation, the presence of permafrost, and geothermal activity within a dry terrain. The interaction between limited snow events and wind results in snow patches buried under a dry soil surface. This leads to ephemeral water streams that only flow during daytime hours. On this volcano, which has the highest located subsurface temperature monitoring systems reported to date, seasonal melting of the permafrost is followed by fast percolation events. This is due to the high porosity of these soils. The results are landforms that shaped by the strong winds. At this site, both thermal springs and lakes (the latter arising from melting ice) provide habitats for life; a 6480m high lake heated by volcanic activity shows both warm and cold sediments that contain a number of different microbial species, including psychrophiles. Where the permafrost melts, thawing ponds have formed at 5900m that is dominated by populations of Bacteroidetes and Proteobacteria, while in the pond sediments and the permafrost itself Acidobacteria, Actinobacteria, Bacteroidetes, Patescibacteria, Proteobacteria, and Verrucomicrobia are abundant. In turn, fumaroles show the presence of acidophilic iron-oxidizers and iron-reducing species. In spite of the extreme conditions reported at Ojos del Salado, this site is easily accessible.

# 1. Introduction

This special issue of *Astrobiology* covers the results of analog research that took place in 2018 at a Mars relevant area on Earth, the Atacama-Altiplano region. This is a wider Atacama Desert area, at the rarely visited Ojos del Salado volcano. The term "site" in this work is used for areas that are several 10s or some 100 km in diameter and show similar Mars-relevant characteristics. Therefore, these site areas could be considered, in aggregate, a "single site" that covers a large area. The site hosts the highest volcano on Earth, where desiccated lakebeds, hot springs, volcanic alluvium within the combined occurrence of permafrost, ephemeral snow, andextreme UV irradiation occur. This terrain is eroded and was once glaciated but is currently desiccated and strongly wind-chilled making it an interesting target for astrobiological research. The major findings are summarized briefly below and are grouped into main topics according to corresponding papers in this special issue. A comparison with other Mars analog sites is presented in the Discussion section.

**Ojos del Salado** (6893 m, 27º06'34" S, 68º32'32" W) is the highest volcano on Earth (Óscar, 1995), with an elevation of 6893 m, located in the Atacama-desert – Altiplano region of the Dry Andes, above the Puna de Atacama plain (Figure 1). The volcano probably represents a Pliocene-Quarternary volcanic arc in the Andes (Rubiolo and Hickson, 2002). Its activity began at around 26 Ma

(million years) ago (Mpodozis et al., 1996) and produced rocks mostly of evolved geochemistry and complex petrography (DeSilva and Francis, 1991) with a recent shift toward more acidic compositions (Bakero et al., 1987). The last volcanically active period was around 30 kyr ago (Moreno and Gibbons, 2007); currently, only fumarolic activity is observed. This volcano represents the latest, and possibly currently active, episode of volcanism from the Miocene until the Pliocene (Goss, 2008) with variations related to Nazca–South American plate convergence parameters and the consequence of subduction of the Juan Fernández Ridge hotspot track (Kay and Mpodozis, 2002). The region is a desert without continuous vegetation or soil cover. Despite the weak and rare precipitation at the closed basins, several salty lakes or their dried up deposits can be found there. Due to the undulating topography and high-altitude permafrost is present (Ahumada, 2002; Cobos and Corte, 1990) with some glacier remnants (Oyarzun, 1987). The morphology of the surface is generally dominated by rocks and boulders, although at a few locations ripples (rarely dunes) are also observable.

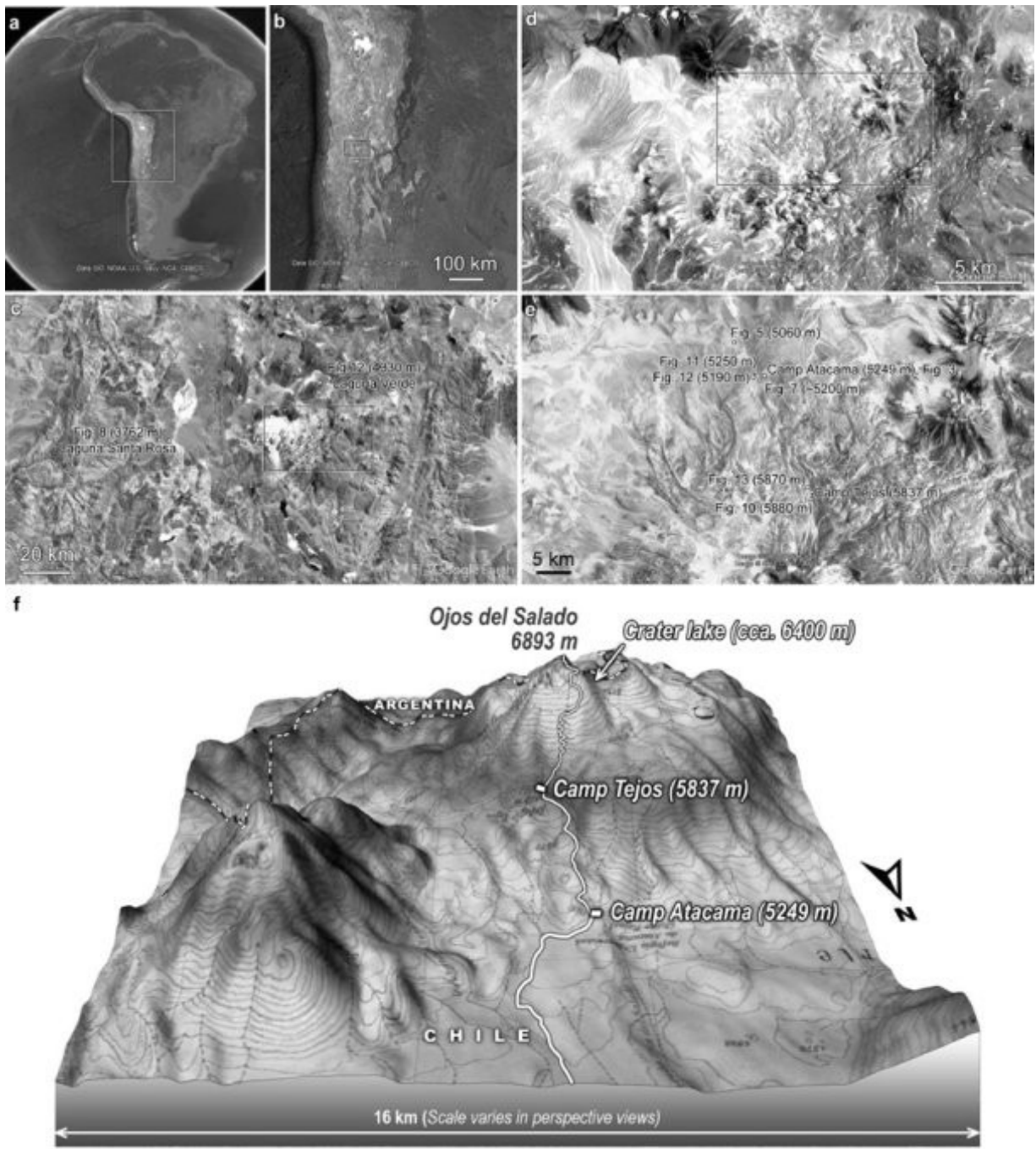

Figure 1. General overview of the Ojos del Salado region based on GoogleMaps images. a: Southern-America, b: the meridional range of the Andes Mountain, c: part of the Altiplano - Atacama Desert region, d: the Ojos del Salado region, e: the northern slopes of the target mountain, f: perspective view of the target area. North is upward in all insets, except the last one (f). The locations of the images in the following figures are indicated in 'c' and 'e' panels.

Ojos del Salado has a high altitude snowline (no measured data exists as this or higher elevations at this site, Clapperton, 1994); however, above 6500 m the terrain is mostly, though not always, snow covered (Amman et al., 2001) and has the highest altitude lakes and desert region on Earth. The **climatic**

**conditions** are only moderately known, partly because of the stochastic changes and the poor coverage with observations; however, general dryness and coldness are expected.

There are no glaciers in the region at the present; however, there were several active glaciers (Nagy et al., 2019) during the Last Glacial Maximum about 19 ka (thousand years) (Ammann et al., 2001). A range of regolith **covered ice and sand masses** occur in various geological settings. **Permafrost** is present in the form of patches above 5200 m and continuously above 5600 m. Degradation of permafrost is an important water source, where, based on model calculations, gradually increased temperature is expected for the next decades. The climatic change, however, will probably not turn this region more humid. The annual precipitation is below 150 mm/year, but it is difficult to obtain an exact measure as there have been only scattered observations made (for more details see the accompanying paper of Breuer et al., 2020 in this Spec. Coll.). Long-term cryosphere monitoring started in 2012 with the installation of loggers at 5 different altitudes. Monitoring has been ongoing ever since; specific details of related analyses is presented by Nagy et al. (2020) in this issue.

The interaction between rare snow events and strong wind-transported sediments produces buried snow masses that are shielded from sublimation. This joint occurrence is rare among Mars analog sites. Even though melting could arise occasionally, the modification of the surface is dominated by desiccation of the cryosphere and by wind activity. The ephemeral surface runoff produces evaporitic sediments that are composed of carbonate, aragonite, anhydrite, talc, and clay minerals, all partly desiccated after deposition. The UV-induced surface modification is strongest here in comparison to other Mars analog sites. A favorable logistical characteristic of this site is its accessibility, with moderately low transport costs, as it is reachable up to 5200 m altitude with a 4-5 hour drive by four wheel drive supported pick-up cars from larger cities such as Copiapó.

## 2. Key Results Presented in this Special Collection of Papers

Specific climatic characteristics of the Ojos del Salado volcano area are described in the paper "**Analog site experiment: surface energy budget components on Ojos del Salado, in the Altiplano-Atacama region (Breuer et al., 2020)**" The meteorological modeling provides new information in at least two aspects. First, it shows examples of how to extrapolate results to larger areas ~ 100 km in diameter by using a few locations where meteorological data have been acquired at the surface. Such work is critical for future Mars exploration, given that surface measurements on the red planet will be available from only a few, scattered locations. The second aspect is modeling at an undulating surface with the kind of overall dryness and reduced atmospheric pressure that typifies the Ojos del Salado region. The handling of meteorological data and forecasting under such conditions might provide useful experience for similar calculations applicable to the even lower atmospheric pressure and dryer conditions of a martian near-surface layer (Weidinger et al., 2009). Remote sensing data for Mars are abundant, though surface in-situ verification is rare. Earth-based methodological improvement thus helps to extrapolate below the surface-based data. The Weather Research and Forecasting (WRF) model used here was verified through surface conditions and energy budget calculations. The results show that near-surface atmospheric conditions can be simulated well, although modeling of the effects of water phase changes should be improved.

Another paper in this special Issue, "**The thermal behavior and surface evolution of ice-bearing ground as an analogue for conditions on Mars: the case of the highest cold, dry desert on Earth, at Ojos del Salado in the Altiplano-Atacama region,**" by Nagy et al. (2020) describes results for monitoring and modeling of continuously changing frozen ground. The temperature variation of the regolith has been monitored hourly since 2012, that is, between 4200 m and 6893 m of altitude, which makes this the highest shallow ground-temperature monitoring system on Earth. The temporal change of water/ice phases was determined, the results of which show that the dryness and cold control the evolution of landforms simultaneously, as is the case for Mars. The presence of permafrost and ground ice was found to be discontinuous above 5200 – 5300 m and continuous above 5800 m. Above 6500 m, the active layer is either missing or extremely thin. The presence of pore ice allows the highest ponds on Earth to exist continuously, though permafrost degradation threatens the extreme wetland habitat.

One of the most interesting possibilities, the potential emergence of liquid water, is discussed in the paper entitled "**Unique and potential Mars relevant flow regime and water sources at a high mountain Altiplano-Atacama site**" by Kereszturi (2020). This work documents the discovery of two types of flow-produced structures and a unique flow regime. Among them, the so-called infilled valleys showed activity only during daytime in a spatially sectioned distribution, originating from three distinct sources: buried snow, surface snow, and ice left behind from liquid water that emerged the day before. Ojos del Salado is a unique ecological site, where material from rare snow events may be protected by the burial of strong wind-transported sediments, which also support ephemeral melting over the long term.

The paper titled "**Effects of Active Volcanism on Bacterial Communities in the Highest-Altitude Crater Lake of Ojos del Salado**" **by Aszalós et al. (2020a)** provides examples for locations of extremophiles at the highest altitude lake (6480 m) on Earth. This lake is heated by geothermal activity, where acidic conditions are produced by fumaroles. Cultivation of sampled organisms showed the presence of psychrophilic taxa that belong to Proteobacteria, Bacteroidetes, and Actinobacteria. Characteristically different bacterial communities were identified in the warm and cold water-covered sediments, with acidophilic iron-oxidizers and iron-reducers in the fumaroles' runoff channel or creek.

Further analysis of these extreme habitats is presented in the paper titled "**Bacterial diversity of a high-altitude permafrost thaw pond located on the Ojos del Salado volcano, Dry-Andes**" by Aszalós et al. (2020b), which characterizes extremophiles in permafrost degradation-produced thaw ponds at 5900 m altitude. Water from the thaw pond was dominated by Bacteroidetes and Proteobacteria, while in the sediment of the lake and permafrost, members of Acidobacteria, Actinobacteria, Bacteroidetes, Patescibacteria, Proteobacteria, and Verrucomicrobia groups were abundant. The lower altitude part of Atacama sites also provides insight, in general, on the methods and problems encountered during laboratory analysis of extremophiles.

The paper titled "**Biosignature Analysis of Mars Soil Analogs: Challenges and Implications for future Missions to Mars**" by Aerts et al. (2020) examines biological fingerprints at several locations at the elevated Andean highlands. It was revealed that although the amino acid load, organic carbon, and nitrogen quantities were generally low, most of the soil samples harbored complex microbial communities dominated by halophilic microorganisms. While the Atacama Desert and higher altitude sites, such as the Ojos de Salado volcano, contain some of the driest, harshest, and highest-altitude environments on Earth, a singular lesson from their study is that, regardless of these kinds of conditions, life is able to survive and potentially be preserved. Salars and salt flats can, therefore, be considered targets of interest.

## 3. Comparison with other Mars analog sites

Collectively, the papers in this special Issue of *Astrobiology* contribute to a better understanding of the geology and astrobiology of Mars through a systematic comparison (Table 1) of the Ojos del Salado site with other potential Mars analog locations on Earth, including many of those previously studied (Hipkin et al., 2014).

Table 1. Comparison of Mars analog sites on Earth to better understand geology (Foucher et al., 2014) and astrobiology (Westall et al. 2013) of Mars. The low and middle altitude Atacama sites cover all other previously analyzed sites except Ojos del Salado as discussed in this special issue. Acronyms: HMP – Haughton Mars Project, FMARS – Flashline Mars Analog Station, MDRS – Mars Desert Research Station. This table does not contain submarine (Thirsk et al., 2007), cave sites (Boston et al., 2001), the sites for analog research of other bodies besides Mars, and the sites used for technological testing (Ross et al., 2013).

| name, location | general characteristics | research topics | logistical aspects | references |
|---|---|---|---|---|
| Antarctic McMurdo Dry Valleys | several sites at a 10-100 km distance from each other | permafrost, cold and dry surface evolution, ephemeral flow features, wind action, brine ponds | very expensive site, transport by flight then by special cars | Marchant & Head, 2007; Tamppari et al., 2012 |
| Svalbard (Norway) | chain of sites at 100-200 km distance from each other | permafrost, ephemeral flow features, cold surface evolution | moderately expensive, reachable by ships and on foot | Hausrath et al., 2008, Bernhardt et al., 2017 |
| Morocco, south from the Atlas Mountain | chain of sites at some 100 km sized area | surface evolution at desert area, ephemeral water covered plains | regular motorway access | Ori et al., 2015, Oberlin et al., 2018 |
| Rio Tinto, (Spain) | sites with locations within 10 km distance | Fe-based biogeochemical system, subsurface | regular motorway access | Amils et al., 2007; Mavris et al., 2018 |

| | | biochemistry | | |
|---|---|---|---|---|
| HMP and FMARS (Devon Island, Canada) | sites with locations within 10 km distance | permafrost, impact crater analysis and related hydrothermal system | moderately expensive, all equipment has to be carried by private flights from Resolute Island | Lee et al., 2007; Binsted et al., 2010; Grau et al., 2018 |
| MDRS (Utah, USA) | locations within 10 km distance | manned mission simulation, extremophiles, technology of field work | regular motorway access, 10 km off-road drive | Direito et al., 2011; Martins et al., 2011 |
| Mauna Kea (Hawaii) | locations within 10 km distance | manned mission simulation, mineralogy, geochemistry | regular motorway access, short off-road drive | Graff et al., 2013; Yingst et al., 2015 |
| Iceland | locations within 100 km distance | ice-volcanism interaction, microbes inside ice, basaltic terrains | regular motorway access | Hartmann et al., 2003; Bathgate et al., 2015 |
| Atacama Desert low and middle altitude (Chile) | chain of sites within some 100 km | extreme desiccation surface evolution, evaporates, salt tolerant organisms | in most cases regular motorway access, then a short off-road drive | Morgan et al., 2014; de Silva et al., 2013; Smith et al., 2014 |
| Ojos del Salado (Altiplano, Chile) | locations within some 10 km distances | permafrost, ephemeral flow features, UV radiation, high altitude geothermal lakes, evaporates | regular motorway access, then a 20 km off-road drive | Aszalós et al. 2020a; 2019b, Breuer et al., 2020 |

Based on Table 1, the Ojos del Salado site is unique in two main aspects. First, there is high UV irradiation, which is more intense here than at any other currently used analog site. It is possibly higher than other potential Altiplano sites such as Llullaillaco (elevation 6723 m), Licancabur (5920 m), and Socompo (6051 m) volcanoes. The second is the presence of geothermal activity and permafrost within a surface terrain that is cold and dry. On Iceland, volcano-ice interactions occur with the surface ice under a cold, though humid, climate, where liquid water produced by melting could be moderately stable. This provides interesting targets, though without the presence of permafrost and an overall dryness, such an environment is a bit less Mars relevant than Ojos del Salado. The Antarctic Dry Valleys site also provides low temperature and permafrost, but without much geothermal activity.

Ojos del Salado site is more Mars relevant than most sites due to the joint occurrence of ephemeral snow precipitation, extreme dryness, low temperatures coupled with strong winds, and easily transportable, low density volcanic grains. This setting provides ideal conditions for the fast burial and maintenance of snow masses under dry and cold conditions.
These serve as a potential analog for the formation of the martian latitude-dependent mantle (LDM), the maindfsfr3 ice-containing layer on Mars (Schon et al., 2009). The strong wind and easily transportable material are necessary to make such interactions possible. These conditions are absent at other Mars analog sites, with the possible exception of the Antarctic Dry Valleys – surface debris may be less erodible there, however, than at the Ojos de Salado. The existence of permafrost at a dry terrain is rare in itself, and therefore only the Antarctic Dry Valleys could be considered an analog site with similar conditions in this aspect.

## 4. Conclusion: the unique characteristics of Ojos del Salado

The papers in this special issue demonstrate that the Ojos del Salado region in the Altiplano and larger Atacama Desert area provides several additional analogue environments that are comparable with other Mars analog sites on Earth. The unique characteristics of the Ojos del Salado site include strong UV irradiation, permafrost in dry terrains, geothermal heat/activity combined with the presence of permafrost, and a strong interaction between ephemeral snow and wind activity. One of the main benefits of this site, when compared to other extreme sites with such dryness and low temperature conditions, is the low cost and simple site access by pick-up trucks. This unique setting provides ideal conditions for the analysis of weathering in dry terrains, permafrost, snow burial, and related geomorphological aspects.

The extremophiles identified and described in this special issue provide insight into the characteristics of microbial communities in dry and cold, geothermally heated terrains, together with the potential to test technological and observational aspects. This special issue of *Astrobiology* presents insight into the potential Mars analog site, Ojos del Salado, and those unique research opportunities that the region has to offer.

## 5. Acknowledgment

The research work realized at the Ojos del Salado region was supported by the following projects and funds: The travel to the area was supported by the COOP_NN_116927 project, the drilling was supported by the ESA EXODRILTECH project and the subsequent laboratory analysis of cosmic analogue material by the GINOP-2.3.2-15-2016-00003 project of NKFIH, and the support from NKFIH for the Size and Shape Laboratory. Part of the field technology and facilities supported by the Hungarian Astronomical Non-profit Ltd. The logistical support is acknowledged from the *Földgömb az Expedíciós Kutatásért Alapítvány*, from the *Földgömb Magazine*, and also from the helpful staff members of the Embassy of Hungary in Santiago, especially from Ambassador Ms. Verónica Chachin (Embassy of Chile in Hungary) and Ambassador Miklós Deák (Embassy of Hungary in Santiago de Chile).The finalization of the manuscript was supported by Szabados L. and Willinger G.M.